\documentclass[a4paper,twoside,twocolumn,english,aps,floatfix,superscriptaddress,showpacs]{revtex4}
\usepackage{times}
\usepackage[T1]{fontenc}
\usepackage[latin1]{inputenc}
\usepackage{amsmath}
\usepackage{graphicx}
\usepackage{amssymb}
\usepackage{babel}


\usepackage{babel}
\makeatother
\begin{document}

\title{Quantum dynamics of two bosons in an anharmonic trap: Collective vs internal excitations}

\author{Christian Matthies}


\affiliation{Theoretische Chemie, Physikalisch-Chemisches Institut, Universit\"{a}t
Heidelberg, INF 229, 69120 Heidelberg, Germany}

\author{Sascha Z\"{o}llner}

\email{sascha.zoellner@pci.uni-heidelberg.de}

\affiliation{Theoretische Chemie, Physikalisch-Chemisches Institut, Universit\"{a}t
Heidelberg, INF 229, 69120 Heidelberg, Germany}

\author{Hans-Dieter Meyer}


\affiliation{Theoretische Chemie, Physikalisch-Chemisches Institut, Universit\"{a}t
Heidelberg, INF 229, 69120 Heidelberg, Germany}

\author{Peter Schmelcher}

\email{peter.schmelcher@pci.uni-heidelberg.de}

\affiliation{Theoretische Chemie, Physikalisch-Chemisches Institut, Universit\"{a}t
Heidelberg, INF 229, 69120 Heidelberg, Germany}

\affiliation{Physikalisches Institut, Universit\"{a}t Heidelberg, Philosophenweg
12, 69120 Heidelberg, Germany}

\date{\today{}}

\begin{abstract}
This work deals with the effects of an anharmonic trap on an interacting
two-boson system in one dimension. Our primary focus is on the role
of the induced coupling between the center of mass and the relative
motion as both anharmonicity and the (repulsive) interaction strength
are varied. The ground state reveals a strong localization in the
relative coordinate, counteracting the tendency to fragment for stronger
repulsion. To explore the quantum dynamics, we study the system's response
upon (i) exciting the harmonic ground state by continuously switching
on an additional anharmonicity, and (ii) displacing the center of mass, this way triggering collective oscillations.
The interplay between collective and internal dynamics materializes
in the collapse of oscillations, which are explained in terms of few-mode
models.
\end{abstract}


\maketitle

\section{Introduction}

Ultracold bosonic atoms have become a popular tool for simulating
and understanding fundamental quantum phenomena (see Refs.~\cite{pitaevskii,pethick,dalfovo99,leggett01}
for an overview). The key reason is that, drawing on their
interaction with electromagnetic fields, cold atoms permit a tremendous
degree of tunability of both external (trapping) forces and atomic interactions.
Typically, however, studies of these kinds of systems rest on two
essential premises: harmonicity of the trap, and mean-field interactions.
The \emph{}first assumption is in fact very natural and easily fulfilled
experimentally, at least near a well-defined minimum of the potential.
In fact, the separability of the center-of-mass motion for identical
particles facilitates many theoretical investigations and has important
ramifications for the dynamics, such as the existence of undamped
dipole oscillation. The obvious question as to how additional anharmonic
forces influence the eigenstates as well as the dynamics has attracted
little attention. In ref. \cite{deuretzbacher2007}, e.g., the effect
of anharmonic terms on the spectrum of two unequal atoms is studied,
whereas Ref.~\cite{li2006} addresses the collapse of collective
oscillations, but from a mean-field perspective. 

Within that mean-field approximation, all atoms are assumed to coherently
populate a single-particle orbital, which is governed by the Gross-Pitaevskii
equation \cite{pitaevskii}. While this is justified in the limit
of atom numbers $N\gg1$ and sufficiently weak interactions, it neglects
two-body effects that become relevant for few-body systems, and in
particular in the limit of strong transversal confinement, when the
system becomes effectively one-dimensional (1D). In that case, it
has been shown \cite{Olshanii1998a} that the effective coupling strength
can be tuned at will, making it possible to change from a non-interacting
to a strongly repulsive system. In that highly correlated borderline
case, the bosons are said to fermionize \cite{girardeau60} as they
acquire partly counterintuitive properties similar to an ideal gas
of fermions. This, among other rationales, has sparked many efforts
toward understanding few-boson systems \cite{busch1998,blume02,alon05,deuretzbacher06,streltsov06,zoellner06a,zoellner06b,zoellner07a}. 

The subject of this paper  is to carry out a systematic \emph{ab initio}
analysis of anharmonic effects on the ground state as well as on the quantum dynamics 
of interacting bosons. We focus on the fundamental two-atom case, which already
reveals a rich variety of features, thus paving the way to systems with
more atoms. It is solved utilizing an exact and flexible multi-configurational
approach, which has proven very fruitful in previous studies on the
fermionization of ensembles with up to six atoms \cite{zoellner06a,zoellner06b,zoellner07a}.
Here the ground state is the natural starting point for an investigation
into the nature of the coupling between the center of mass and the
relative coordinate. Equipped with this insight, we probe the impact 
of that coupling on the dynamics via two general schemes: First, we switch
on an additional anharmonic trapping force, this way exciting the
(uncoupled) harmonic ground state. Second, we excite the collective
motion by elongating the center of mass and follow the induced dynamics
in the relative motion. 

This paper is organized as follows: After a brief introduction of
the theoretical model in Section~\ref{sec:theory}, we present our
computational method in Sec.~\ref{sec:compmethod}. The subsequent
section~\ref{sec:gs} deals with the ground state
of the two-boson system in an anharmonic trap. In Sec. \ref{sec:dynamics},
we study the dynamics after excitation.


\section{Model}
\label{sec:theory}


In this work we investigate a system of two interacting bosons under external confinement. 
These particles, representing atoms, are taken to be one-dimensional (1D). More precisely, after integrating out the transverse degrees of freedom and rescaling to dimensionless units, we arrive at the model Hamiltonian (see \cite{zoellner06a} for details) 
\begin{equation}
H = h_1 + h_2 + V(x_1 - x_2),
\end{equation}
where $h_i = \frac{1}{2}p_i^2 + U(x_i)$ is the one-particle Hamiltonian with a trapping potential $U$, while $V$ is the effective two-particle interaction \cite{Olshanii1998a} $V(x) = g \delta_{\sigma}(x)$. Here we concentrate on repulsive foces, $g>0$.
The well-known numerical difficulties due to the spurious short-range behavior of the standard delta-function potential $\delta(x)$ are alleviated by mollifying it with the normalized Gaussian
\begin{equation}
\delta_{\sigma}(x) = \frac{1}{\sqrt{2\pi}\sigma} e^{-x^2/2\sigma^2},
\end{equation}
which tends to $\delta(x)$ as $\sigma \rightarrow 0$ in the distribution sense. We choose a fixed value $\sigma = 0.1$ as a trade-off between smoothness and a range much smaller than the average particle distance.

We assume a harmonic trap superimposed by an anharmonic potential with linear and quartic terms:
$U(x_i) = \frac{1}{2}x_i^2+\kappa x_i +\lambda x_i^4$.
The linear part of the potential causes a displacement of the trap center (controlled by a displacement factor $\kappa$), while the quartic terms squeeze the trap (controlled by an anharmonicity factor $\lambda$). By time-dependent variation of both parameters later on in this work, the trap can be distorted so as to cause excitations in the two-boson system.

For studying the effects of the coupling between the center of mass and the internal relative dynamics, we perform a coordinate transformation to the center-of-mass frame of reference: $R = \frac{1}{2} \left( x_1 + x_2 \right)$ and $x = x_1 - x_2$. 
Finally, we arrive at the following Hamiltonian
\begin{eqnarray}
H & = & -\frac{1}{4} \frac{\partial^2}{\partial R^2} - \frac{\partial^2}{\partial x^2} +
R^2 + \frac{1}{4}x^2 + 2 \kappa R +  2\lambda R^4 + {} \nonumber \\
& &{}  + \frac{\lambda}{8}x^4 + 3 \lambda R^2x^2 + \frac{g}{\sqrt{2\pi}\sigma}e^{-x^2/2\sigma^2}.
\label{eqn:HaRr}
\end{eqnarray}
After the transformation to the center-of-mass frame, the interaction depends only on  the relative coordinate $x$, whereas the shift of the trap only affects the center of mass coordinate $R$. Moreover, for a quartic anharmonicity ($\lambda > 0$) an $Rx$-coupling term appears in the Hamiltonian.

\section{Computational method}
\label{sec:compmethod}

Our goal is to investigate the ground state and the dynamics of the system
introduced in Sec.~\ref{sec:theory} for all relevant interaction
strengths in a \emph{numerically exact,} i.e.\emph{,} controllable
fashion. 
Our approach relies on
the Multi-Configuration Time-Dependent Hartree (MCTDH) method
\cite{bec00:1}, primarily a wave-packet dynamics tool known for its
outstanding efficiency in high-dimensional applications. To be self-contained,
we will provide a concise introduction to this method and how it can
be adapted to our purposes.

The underlying idea of MCTDH is to solve the time-dependent Schrödinger
equation
\begin{equation}
i\dot{\Psi}=H\Psi,\quad\ \Psi|_{t=0}=\Psi^{(0)}\end{equation}
 as an initial-value problem by expansion in terms of direct (or Hartree)
products $\Phi_{J}\equiv\varphi_{j_{1}}^{(1)}\otimes\cdots\otimes\varphi_{j_{N}}^{(N)}$:\begin{equation}
\Psi(\cdot,t)=\sum_{J}A_{J}(t)\Phi_{J}(\cdot,t),\label{eq:mctdh-ansatz}\end{equation}
using a convenient multi-index notation for the configurations, $J = (j_1, \dots, j_f)$, where $f$ denotes the number of degrees of freedom. 
The (unknown) \emph{single-particle functions} $\varphi_{j_{\kappa}}^{(\kappa)}$
($j_{\kappa}=1,\dots,n_{\kappa}$) are in turn represented in a fixed
\emph{primitive} basis implemented on a grid. 
In our work, we consider systems with two degrees of freedom and use $n_R = n_x = 8$ orbitals. 
The grid spacing, in our case, should of course be small enough for the relative coordinate to sample the interaction potential, whereas the center-of-mass coordinate is not that sensitive. In this light, we employ a discrete variable representation with $95$ and $165$ grid points for the center-of-mass coordinate and the relative coordinate, respectively.

Note that in the above expansion, not only the coefficients $A_{J}$
are time dependent, but so are the Hartree products $\Phi_{J}$. Using
the Dirac-Frenkel variational principle, one can derive equations
of motion for both $A_{J},\varphi_{j}$ \cite{bec00:1}. Integrating
this differential-equation system allows one to obtain the time evolution
of the system via (\ref{eq:mctdh-ansatz}). Let us emphasize that
the conceptual complication above offers an enormous advantage: the
basis $\{\Phi_{J}(\cdot,t)\}$ is variationally optimal at each time
$t$. Thus it can be kept fairly small, rendering the procedure very
efficient.


The MCTDH approach \cite{mctdh:package}, which we use, incorporates
a significant extension to the basic concept outlined so far. The
so-called \emph{relaxation} method \cite{kos86:223} provides a way
to not only \emph{propagate} a wave packet, but also to obtain the
lowest \emph{eigenstates} of the system, $\Psi_{0}$. The key idea
is to propagate some wave function $\Psi^{0}$ by the non-unitary
$e^{-H\tau}$ (\emph{propagation in imaginary time}.) As $\tau\to\infty$,
this exponentially damps out any contribution but that stemming from
the true ground state like $e^{-E_{m}\tau}$. In practice, one relies
on a more sophisticated scheme termed \emph{improved relaxation} \cite{mey03:251,meyer06},
which is much more viable especially for excitations. Here $\langle\Psi|H|\Psi\rangle$
is minimized with respect to both the coefficients $A_{J}$ and the
orbitals $\varphi_{j}$. This leads to (i) a self-consistent eigenvalue
problem for $(\langle\Phi_{J}|H|\Phi_{K}\rangle)$, which yields $A_{J}$
as {}`eigenvectors' , and (ii) equations of motion for the orbitals
$\varphi_{j}$ based on certain mean-field Hamiltonians. These are
solved iteratively by first diagonalizing for $A_{J}$ with \emph{fixed}
orbitals and then `optimizing' $\varphi_{j}$ by propagating them
in imaginary time over a short period. That cycle will then be repeated until convergence is achieved.

\section{Ground state of two-boson systems in anharmonic traps}
\label{sec:gs}

\begin{figure}[b]
\centering

\includegraphics[width=6cm]{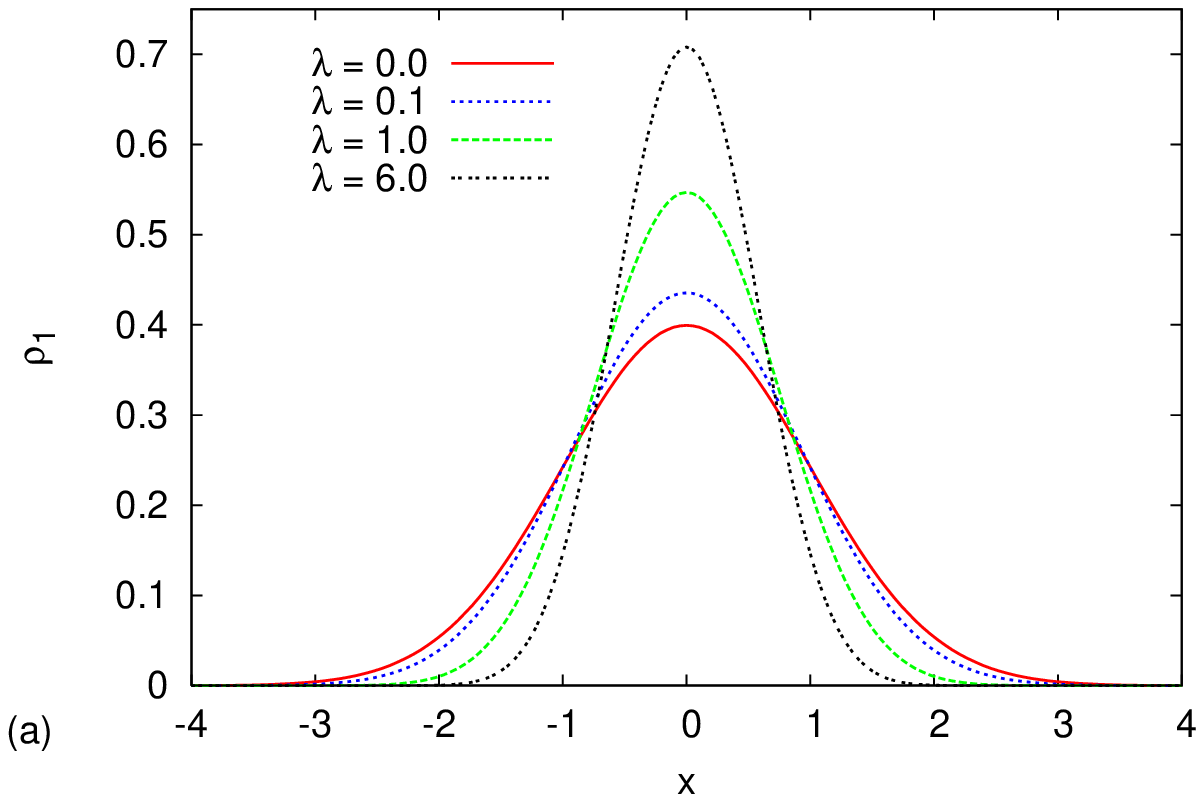} 
\includegraphics[width=6cm]{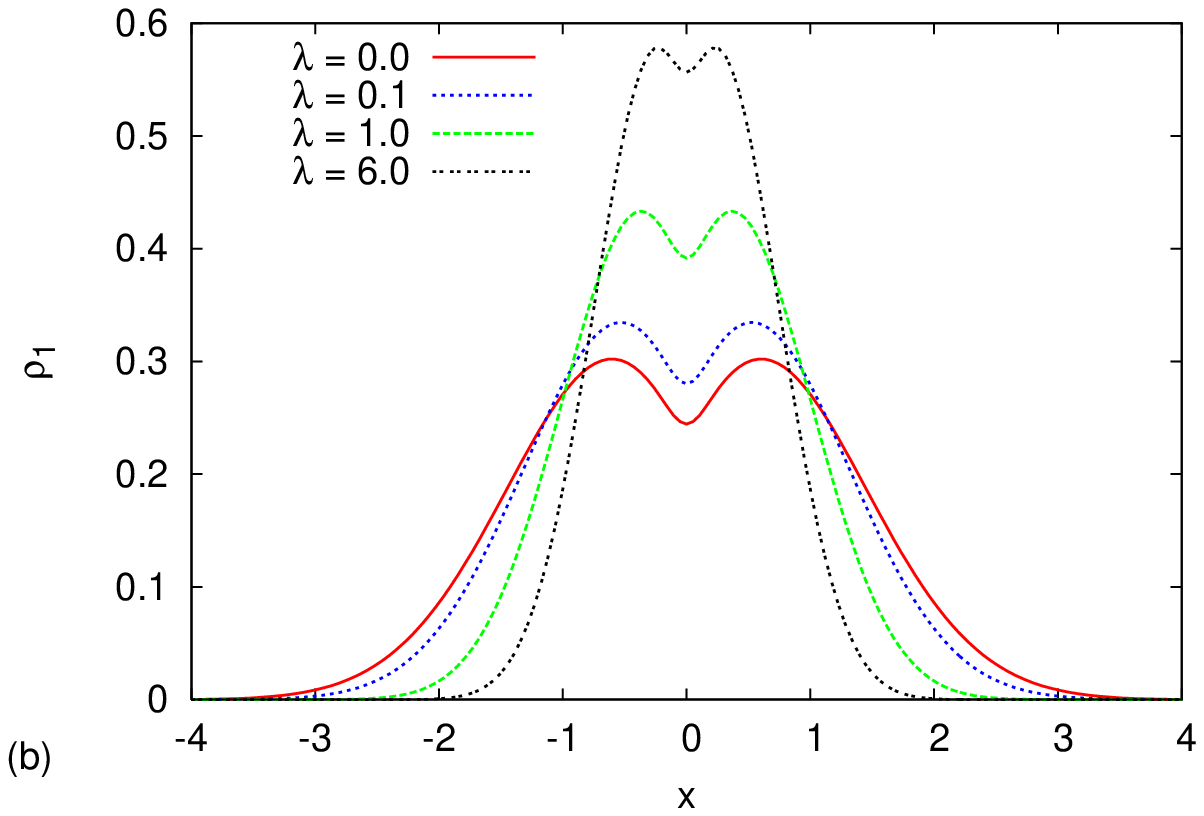} 

\includegraphics[width=6cm]{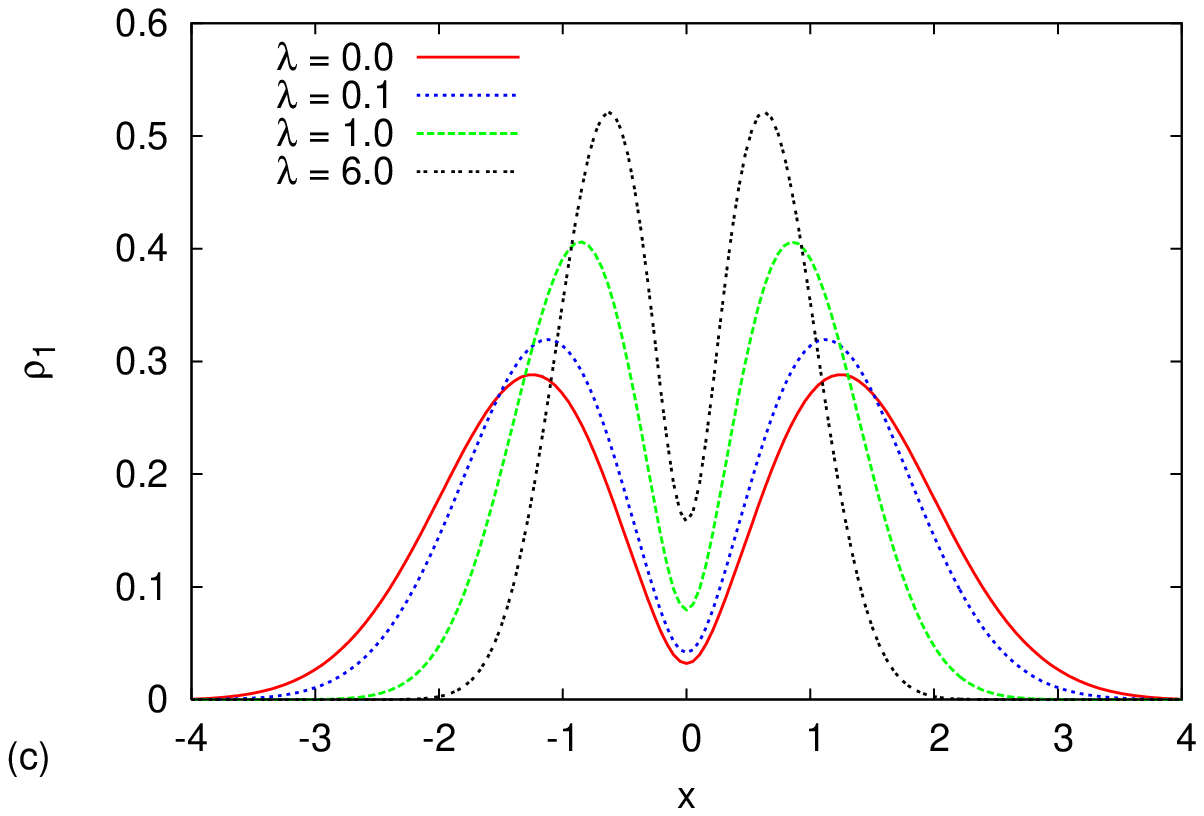} 
\includegraphics[width=6cm]{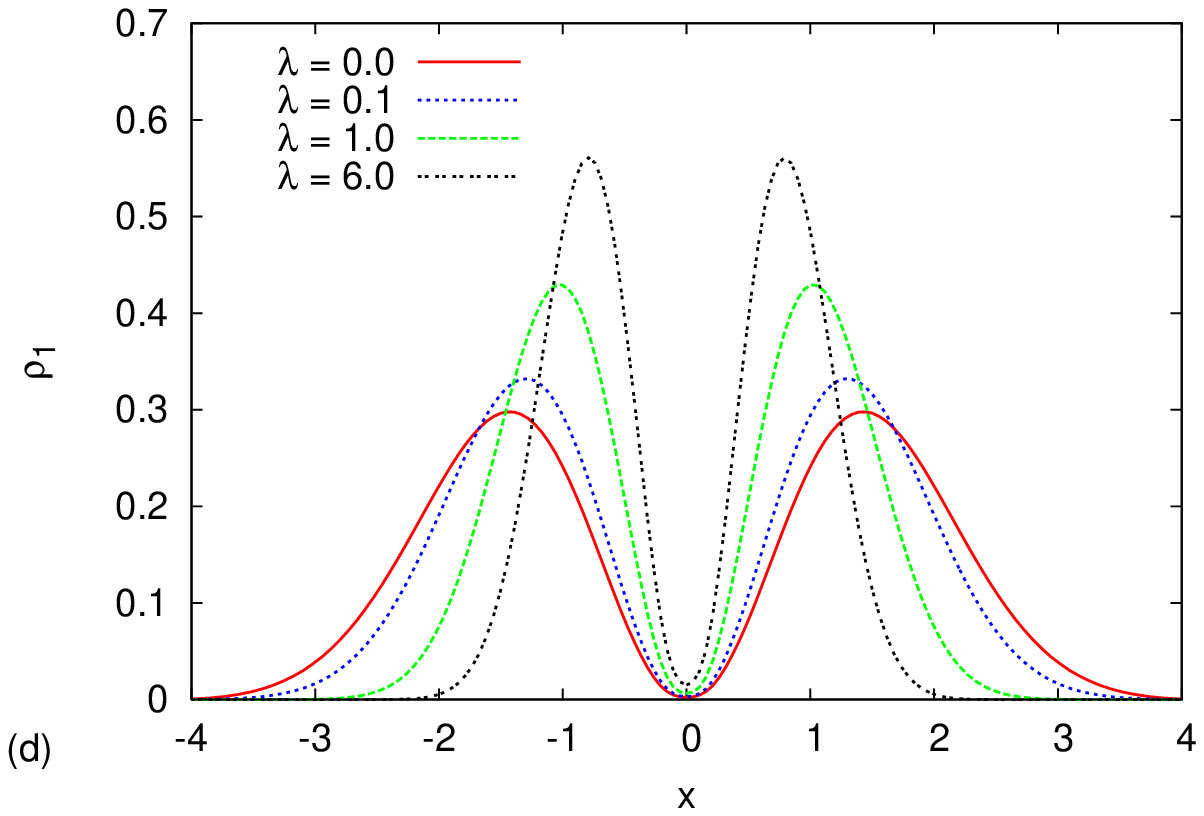} 
\caption{Ground state of the two-boson system in an anharmonic trap with different anharmonicity factors $\lambda$. Shown are the one-particle densities for the relative coordinate  $\rho_1(x)$  for interaction strengths of (a) $g = 0.0$, (b) $g = 1.0$, (c) $g = 6.0$, (d) $g = 20.0$}
\label{fig:d1d} 
\end{figure}

Before studying the quantum dynamics of the interacting two-boson system in an anharmonic trap, we examine its ground state in dependence of the interaction strength $g>0$ and the anharmonicity $\lambda$. 
%
%
By increasing $\lambda$, and thus squeezing the trap, we can force the particles to reduce their distance, this way counteracting the tendency to fragment as the interaction strength $g \rightarrow \infty$.
Figure \ref{fig:d1d} depicts one-particle densities $\rho_1(x) = \int \mathrm{d}R \, |\Psi(R,x)|^2$ of the relative coordinate for different interaction strengths $g$ and anharmonicities $\lambda$ in an undisplaced ($\kappa = 0$) anharmonic trap. Obvioulsy [cf. Fig. \ref{fig:d1d}(a)], for non-interacting particles, the probability density is peaked at $x=0$, i.e., both atoms remain in the trap center regardless of whether there is another one already. For increasing repulsion [Figs. \ref{fig:d1d}(b-d)], the mean distance between the particles grows, and more and more a pronounced fragmentation in the density occurs, indicating the inhibition of both particles being located at the same point in space. At $g = 20.0$ (cf. fig. \ref{fig:d1d} (d)), the fragmentation is almost complete. By contrast, the squeezing of the trap caused by the anharmonic term $\lambda>0$ leads to a reduction of the fragmentation in the density. Both effects make for a stronger localization of the two particles.

\begin{figure}[b]

\includegraphics[width=0.45\columnwidth]{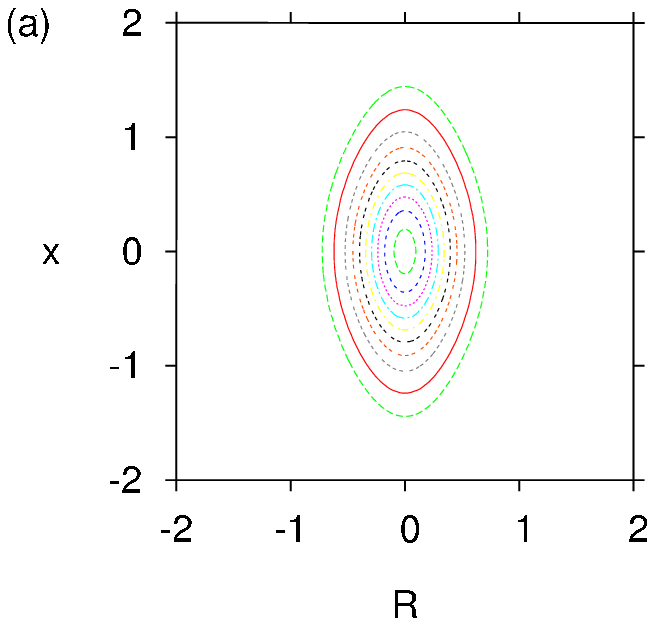} 
\includegraphics[width=0.45\columnwidth]{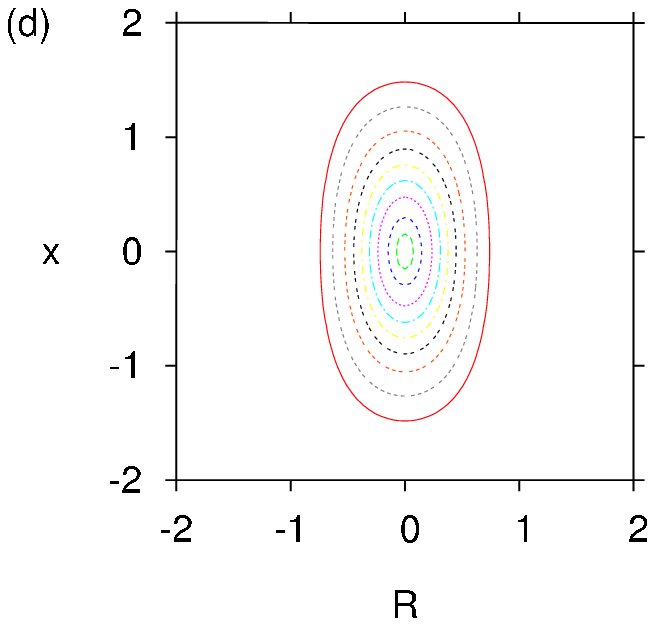} 

\includegraphics[width=0.45\columnwidth]{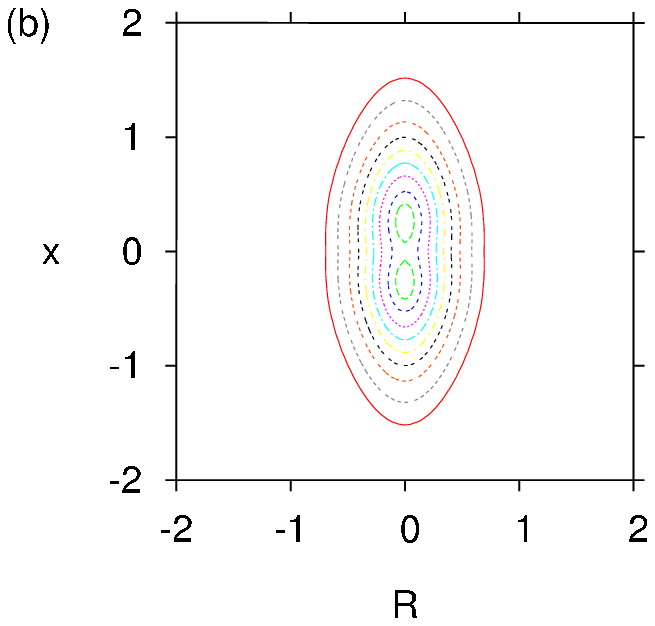} 
\includegraphics[width=0.45\columnwidth]{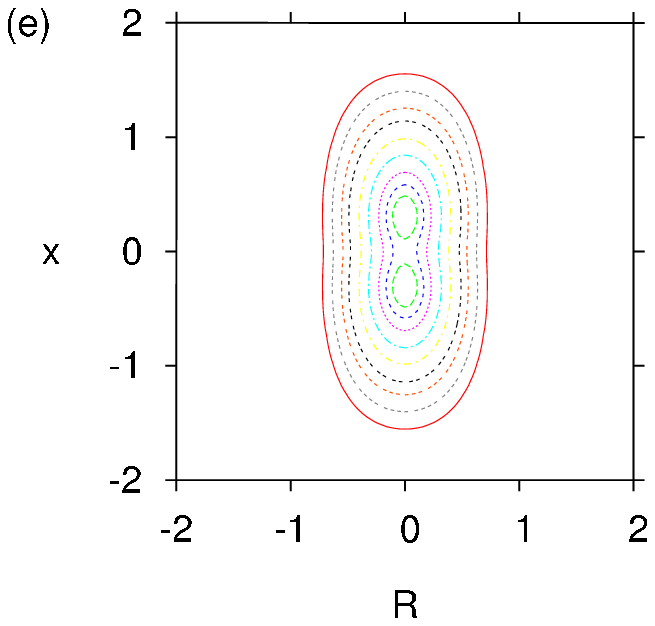} 

\includegraphics[width=0.45\columnwidth]{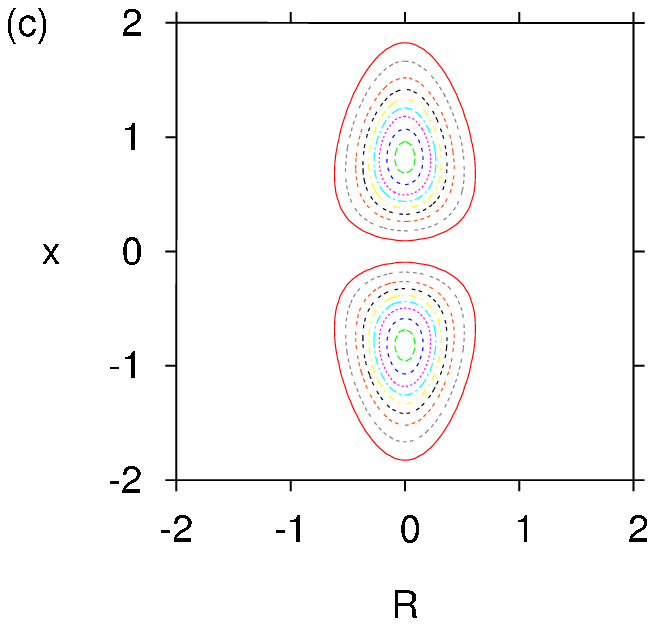} 
\includegraphics[width=0.45\columnwidth]{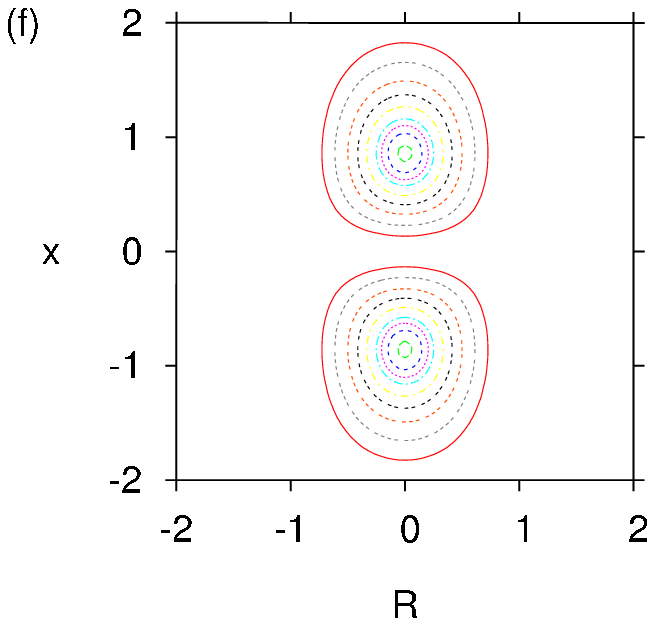} 

\caption{(a) - (c): Two-particle densities $\rho_2(R,x)$ of the ground state of the two-boson system in an anharmonic trap with anharmonicity factor $\lambda = 6.0$ and (a) $g = 0.0$, (b) $g = 1.0$, (c) $g = 20.0$. --- (d) - (f) same, but  without $Rx$-coupling term in the Hamiltonian.}
\label{fig:d2d} 
\end{figure}


Furthermore, to illuminate the role of the coupling between the center of mass ($R$) and the relative coordinate ($x$), we consider the same situation as before,
but we artificially switch off the coupling term $3\lambda R^2 x^2$ in the Hamiltonian in Eq.~(\ref{eqn:HaRr}). 
For a special trap with anharmonicity $\lambda = 6.0$ the corresponding two-particle densities $\rho_2(R,x) = |\Psi(R,x)|^2$ with and without $Rx$-coupling are shown in Fig.~\ref{fig:d2d} for different interaction strengths (note that only for two degrees of freedom, as in our case, the two-particle density is equal to the absolute square of the wave function). 
Comparing the true results [Figs. \ref{fig:d2d}(a--c)] with those where the $Rx$-coupling term has been artificially excluded [Figs. \ref{fig:d2d}(d--f)], it becomes clear just how that coupling suppresses  configurations in which both the center of mass and the relative coordinate are large. In other words, it is even less likely for the atoms be to be far from the trap center and simultaneously far from each other. This is visible in the two-particle densities as an unphysical flattening in the affected regions in the absence of the $Rx$ term [Figs. \ref{fig:d2d}(d--f)], in contrast to the full result. The effect is more pronounced for larger anharmonicities $\lambda$, when the coupling term becomes more relevant (although it is still dominated by the quartic terms like $R^4$ especially for large values of $R$ and $x$).

\begin{figure}[t]
 \centering

\includegraphics[width=0.49\columnwidth]{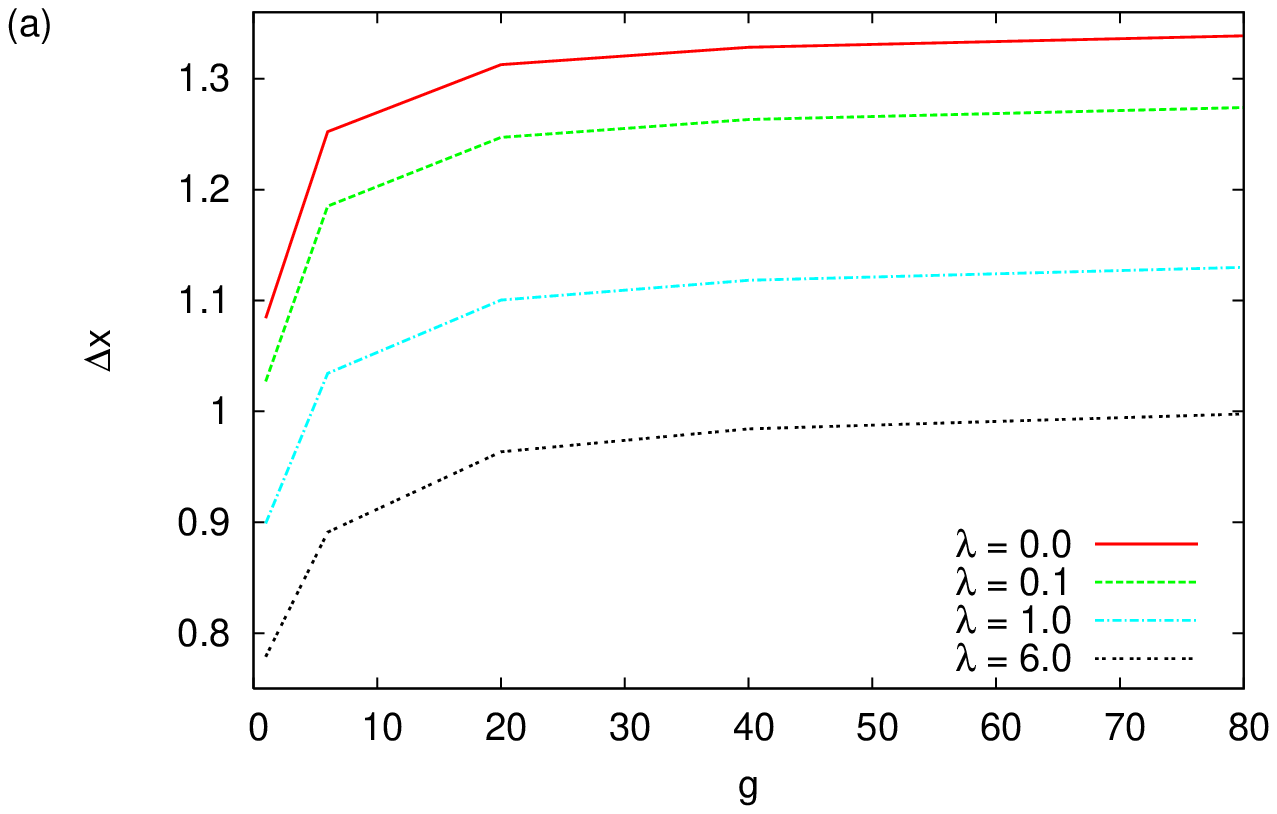} 
\includegraphics[width=0.49\columnwidth]{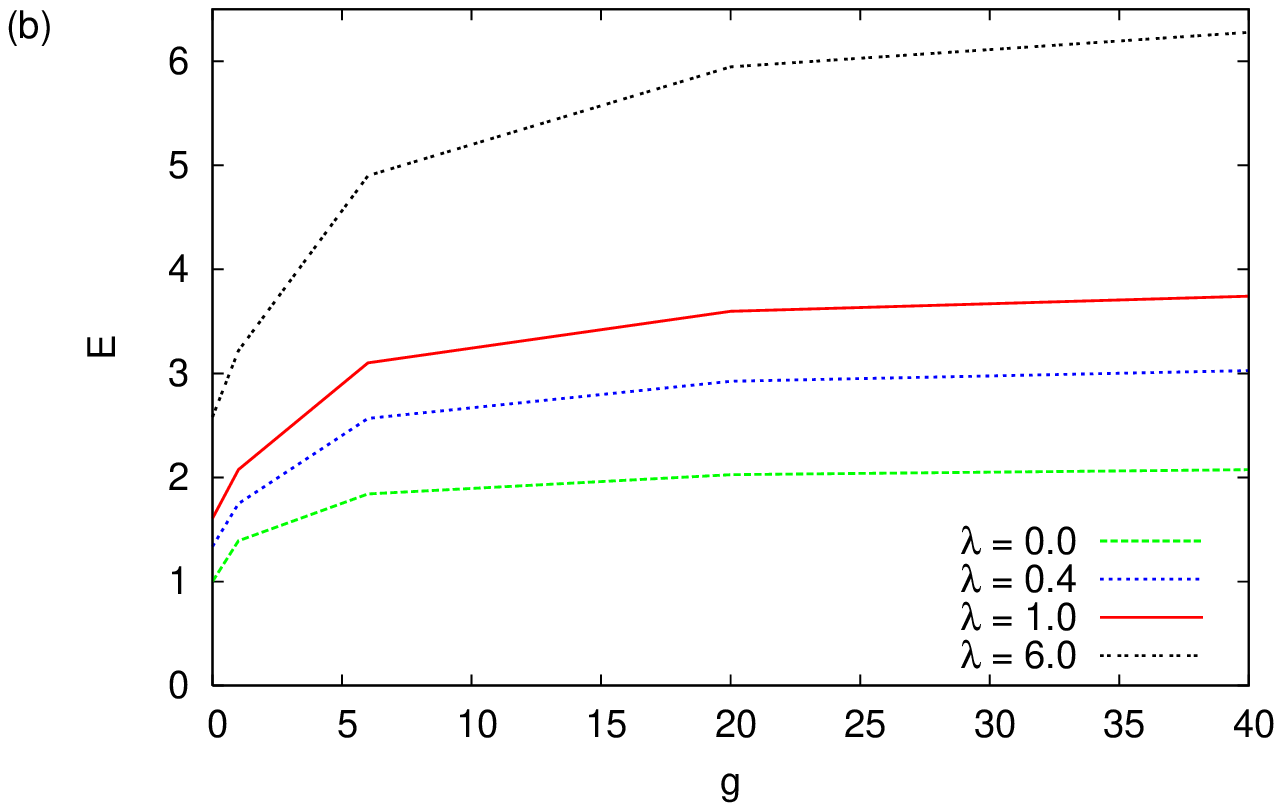} 

\caption{(a) Mean relative distance of the two particles and (b) energy of the groundstate of the two-boson-system both as a function of the anharmonicity factor $\lambda$ and the interaction strength $g$.}
\label{fig:dist_and_ener} 
\end{figure}

To measure the mean relative distance between the two particles, we look at the quantity $\langle x^2 \rangle$, which is equal to the square of the standard deviation $\Delta x$, as $\langle x \rangle=0$ due to permutation  symmetry. It turns out [Fig. \ref{fig:dist_and_ener}(a)] that with increasing interaction strength (for a constant anharmonicity of the trap potential) the mean relative distance first rises quickly and then becomes almost constant because of the limited range of the interaction. By contrast, a larger anharmonicity in the trap potential and thus a stronger confinement of the particles leads to a decrease in the (mean-square) relative distance, as expected. Figure  \ref{fig:dist_and_ener}(b) shows the energy of the ground state as a function of the interaction strength $g$, plotted for different anharmonicity factors $\lambda$. For a constant anharmonicity the energy first increases with growing interaction strength, then more and more saturates because of the finite range of the interaction. For a larger anharmonicity, the energy of the ground state is higher because of the increase in potential energy. Owing to the squeezing, it is harder for the atoms to move apart and isolate each other, this way shifting the fermionization limit to larger $g$.

\section{Quantum dynamics} 
\label{sec:dynamics}

After having explored the ground state, we now want to consider excitations of the interacting two-boson system. Altogether, we apply two methods of excitation, both realized by a distortion of the external trap. In the first scheme, described in Sec.~\ref{sec:probl1}, we switch on a quartic anharmonic potential in addition to a purely harmonic trap; in the other case (Sec.~\ref{sec:probl2}), we displace the trap center by adding a linear potential to a quartic anharmonic trap. In both cases we demonstrate that collapse and revival of oscillations in the relative coordinate occur if the excitation is performed non-adiabatically, that is, if the duration of the switching process is short compared to the relevant time scales of the system. Moreover, we analyze the excitation spectra in order to explain the dynamics within a few-mode model.

\subsection{Switching process}
\label{sec:switch}

Let us first briefly discuss the switching procedure itself. Technically the switching is implemented by multiplying the potential terms to be switched on or off with a specific switching function $\Theta(t)$ which was chosen as
\begin{equation}
\Theta(t) = \frac{1}{2} \left( 1 \pm \tanh \left[\xi(t-\tau)\right] \right).
\label{eqn:switch}
\end{equation}
For all intents and purposes, this function is time independent except for a short while controlled by the parameters $\xi$ (the inverse duration of the switching) and $\tau$ (responsible for the moment in time of switching). The positive (negative) sign corresponds to switching on (off) the potential terms. This way, the Hamiltonian is essentially time-dependent only during the switching process. Before that, the system is in its ground state; after the excitation (for $t-\tau\ge3\gg1/\xi$), it is propagated by the time-independent Hamiltonian (\ref{eqn:switch}), with $\kappa=0$. If the duration of the switching is sufficiently long compared to the time scales of the system, the excitation process occurs adiabatically and the system goes over into to its momentary ground state (for a system with parameters $g = 20.0$ and $\lambda = 0.1$ this is the case for $\xi < 0.5$). In the following we use the values $\xi = 3.0$ and $\tau = 2.0$.

\subsection{Excitation by switching on an anharmonic potential}
\label{sec:probl1}

For the first excitation scheme, we prepare the two-boson system in the ground state of a purely harmonic trap ($\lambda=0$). Then, over a short while characterized by the parameters $\xi$ and $\tau$ (cf. eq. (\ref{eqn:switch})), an additional quartic potential is switched on (with an asymptotic value $\lambda>0$). After that, the system---now again described by the time-independent Hamiltonian (\ref{eqn:HaRr}) ---is propagated for some time $t$, and its quantum dynamics is studied. 


\subsubsection{Collapse and revival of oscillations}

\begin{figure}[t]
\includegraphics[width=0.9\columnwidth]{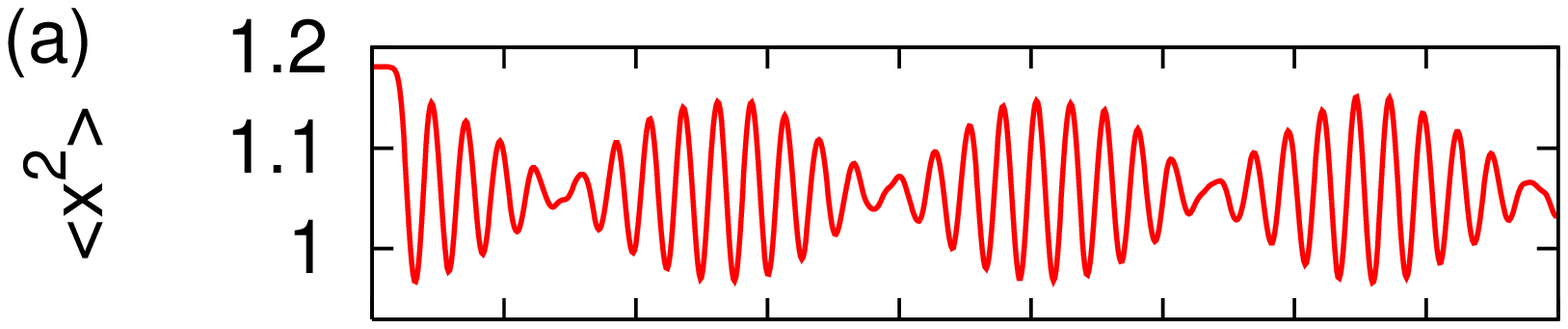}\\
\includegraphics[width=0.9\columnwidth]{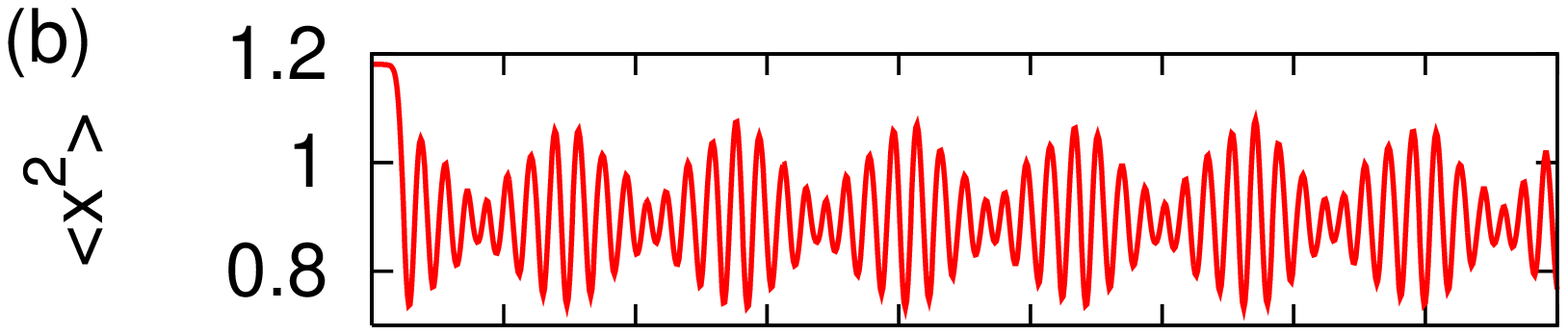}\\
\includegraphics[width=0.9\columnwidth]{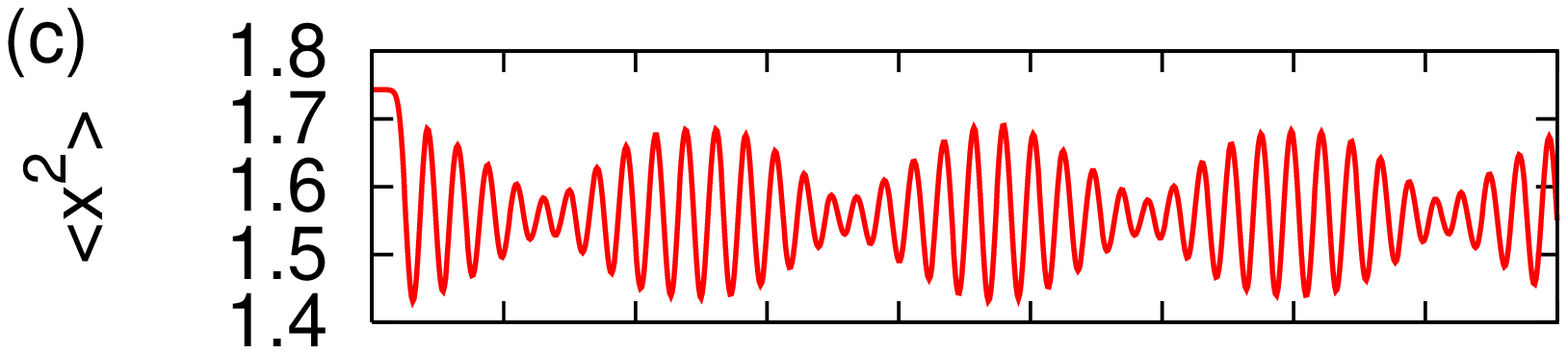}\\
\includegraphics[width=0.9\columnwidth]{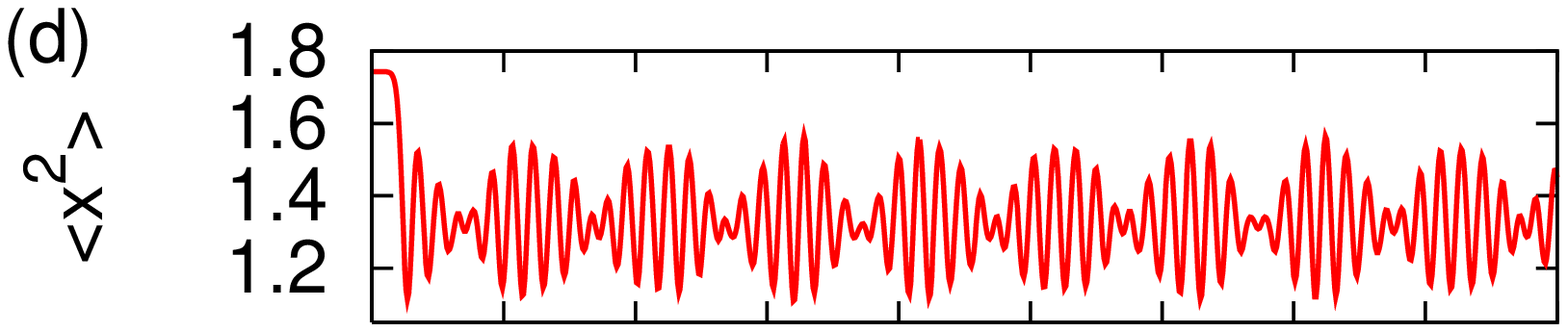}\\
\includegraphics[width=0.9\columnwidth]{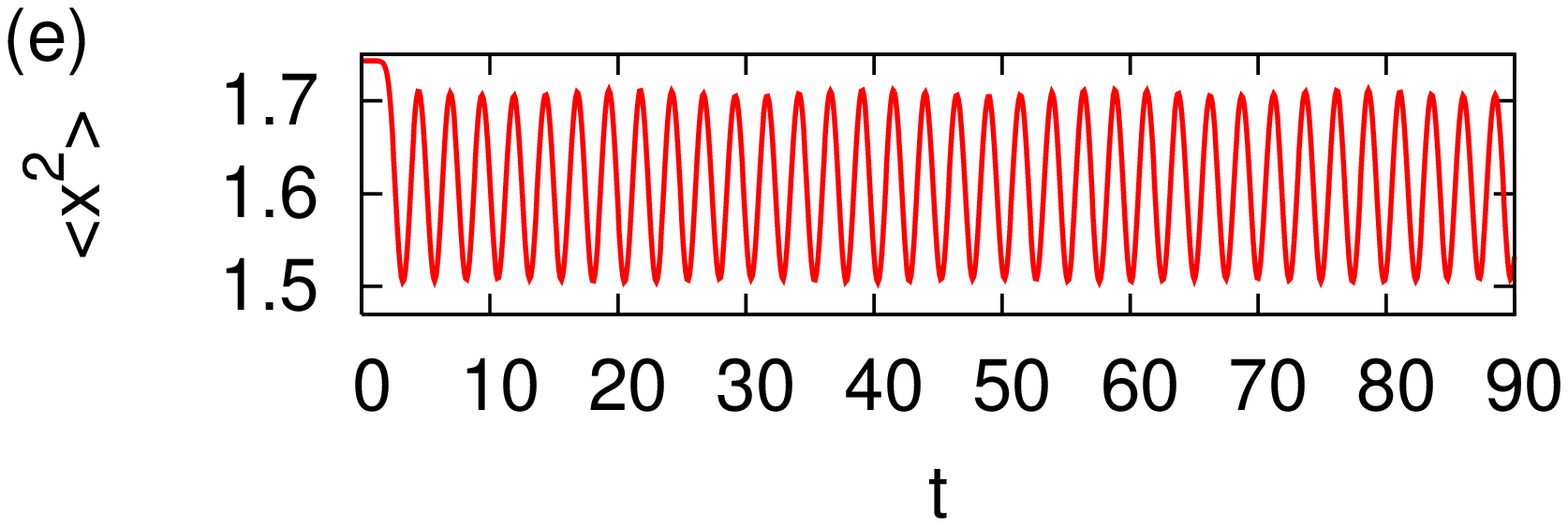}

\caption{Dynamics of the two-boson system after switching on a quartic anharmonicity. Square of the mean relative distance of the two particles $\langle x^2 \rangle$ for (a) $g = 1.0$, $\lambda = 0.1$, (b) $g = 1.0$, $\lambda = 0.5$, (c) $g = 20.0$, $\lambda = 0.1$, (d) $g = 20.0$, $\lambda = 0.5$ and --- (e) $g = 20.0$, $\lambda = 0.1$ (without $Rx$-coupling term)}
\label{fig:collrev} 

\end{figure}

The excitation process outlined above alters the shape of the trap, but without affecting the parity symmetries for both $R$ and $x$; hence, $\langle R \rangle = 0$ for all times (recall that trivially  $\langle x \rangle = 0$ by permutation symmetry alone). However, the squeezing excites breathing oscillations both in the center-of-mass width $\Delta R$ (omitted here) as well as in the internal motion, $(\Delta x)^2=\langle x^2\rangle$, which experiences collapses and revivals
[cf. Figs. \ref{fig:collrev}(a--d)]. There are time spans in which the interatomic distance is rapidly changing, while at other times the system is almost at rest or only oscillating very slowly. The time between two collapses of the dynamics depends on the excitation process, i.e., on  the amount of energy added by the anharmonic potential as well as on the interaction strength between the two particles. It turns out that the stronger the added anharmonic potential (cf. figs. \ref{fig:collrev} (a/b),(c/d)) and the stronger the interaction (cf. figs. \ref{fig:collrev} (a/c),(b/d)) are, the shorter is the time between two collapses. 


\subsubsection{Excitation spectra and symmetry analysis of excited states}


For a better understanding of the effects of the excitation, it is natural to analyze the contributing excited states of the stationary Hamiltonian $H(t\gg1/\xi)$, i.e., including the full anharmonic potential and without the displacement term ($\lambda > 0, \kappa = 0$). We describe the system after excitation by expanding its  wavefunction $\Psi(t)$ in terms of the eigenstates $\Psi_j$ of that  Hamiltonian, 
\begin{equation}
\Psi(R,x,t) = \sum_{j} c_j(t) \Psi_j(R,x),
\label{eqn:psiges}
\end{equation}
where $H \Psi_j(R,x) = \omega_j \Psi_j(R,x)$ and $\omega_0$ denotes the ground state energy. The analysis of the excitation spectrum --- obtained via Fourier transformation of the autocorrelation function $a(t)\equiv \langle \Psi(0)|\Psi(t)\rangle$ --- provides us with the values for {$\omega_j$}. On the other hand, the exact eigenvalues can also be obtained by diagonalizing the Hamiltonian and then be related to the peaks in the spectrum $\tilde a(\omega)$.

\begin{figure}[t]
\includegraphics[width=0.7\columnwidth]{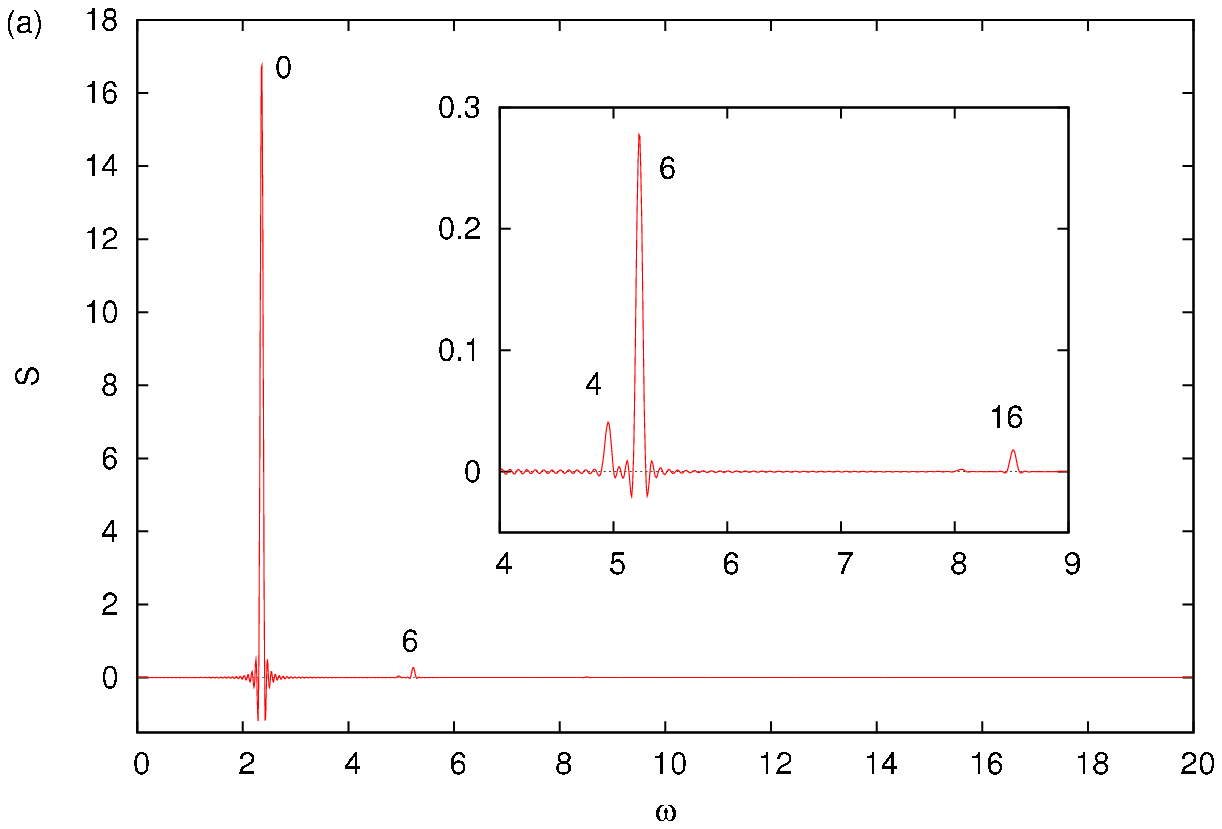} 
\includegraphics[width=0.7\columnwidth]{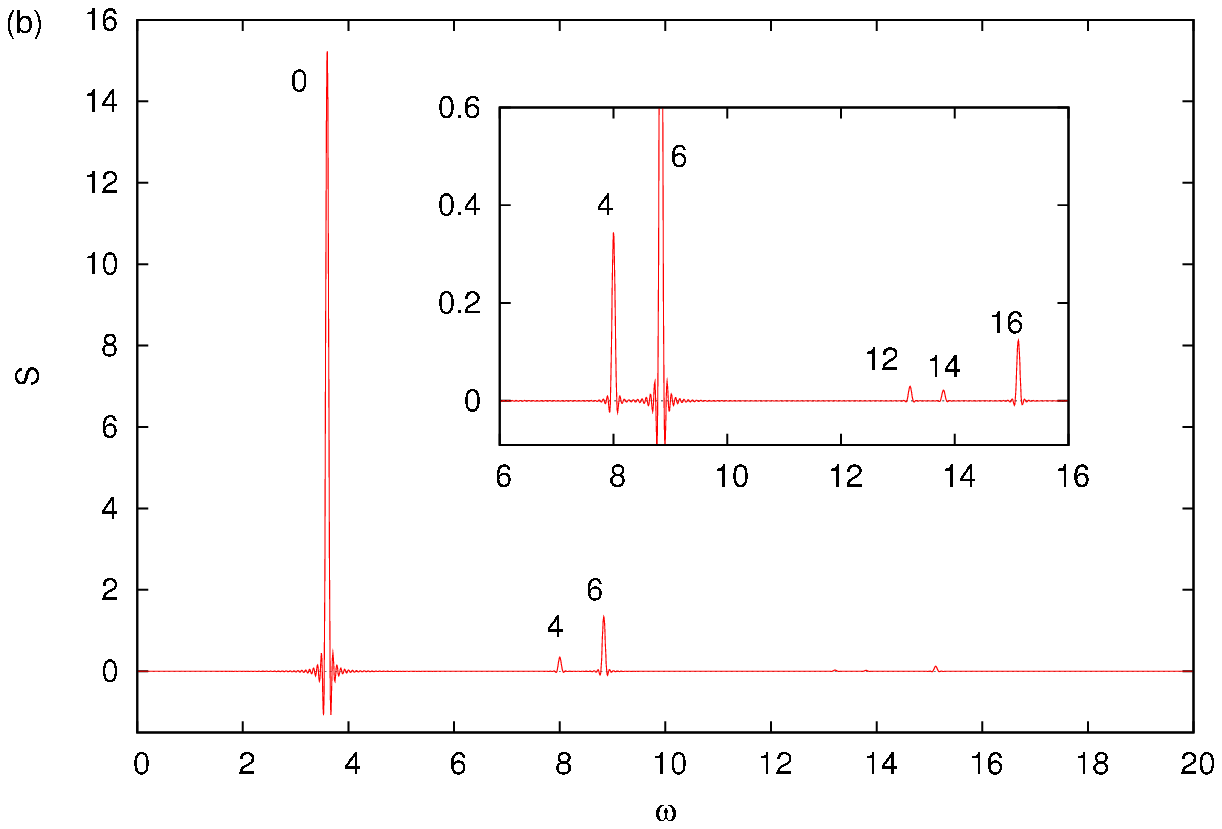} 
\caption{Excitation spectra for (a) $g = 20.0$, $\lambda = 0.1$, -- (b) $g = 20.0$, $\lambda = 1.0$. The numbers indicate the excited states, \textit{0} denotes the ground state.}
\label{fig:spectra}
\end{figure}

This procedure yields that the ground state $\Psi_0$ gives the main contribution (cf. fig. \ref{fig:spectra}) . Furthermore, not all but only certain higher states are populated in (\ref{eqn:psiges}). For very large interaction strengths ($g \gtrsim 10$), apart from the ground state in the anharmonic trap, only the eigenstates $j=4,6,12,14,16$ are excited. For smaller interaction strengths ($g \lesssim 10$), in turn, the lowest eigenstates $j=4,5,12,13,16$ are excited. Generally, higher states are less and less excited while the ground state contains the main contribution. The more energy is put into the system by the distortion of the harmonic trap (cf. fig. \ref{fig:spectra} (a) and (b) for $g = 20.0$ and $\lambda = 0.1$ and $1.0$, respectively) the more strongly energetically higher states are excited. The same is true for fast excitation processes, which likewise deposit more energy into the system. For small distortions of the harmonic trap ($\lambda \lesssim 0.1$) and an adequate duration of switching ($\xi \lesssim 3.0$), only two higher states (the $4^{th}$ and $6^{th}$ for large, the $4^{th}$ and $5^{th}$ for small interaction strengths) are considerably excited while all other states have only a negligible amplitude.

The fact that not all of the lowest eigenstates are excited hints at selection rules due symmetries of the system. Indeed, as is obvious from Eq. \ref{eqn:HaRr} (for $\kappa = 0$), the Hamiltonian conserves both the center-of-mass parity and the relative parity, so only states having the same $R$- and $x$-parities as the initial state can be excited. A closer analysis confirms that only the states mentioned above have even $R$-parity as well as even $x$-parity and thus obey the same symmetries as the initial state, i.e., the harmonic ground state. It should be mentioned that in our way of counting the excited states, we include all states having odd $x$-parity (e. g. the $1^{st}$ or the $3^{rd}$ excited state),  which would represent a non-bosonic system.

\subsubsection{Few-mode model}
\label{sec:fewmode}


The dynamics of the system can be fitted very well to a simple few-mode ansatz. For this, we revert to the expansion  (\ref{eqn:psiges}) of the wavefunction, restricted to $m+1$ eigenstates of the Hamiltonian. Solving the Schrödinger equation fixes the time-dependent coefficients in the most general form as $c_j(t) = c_j \, e^{-i \omega_j t - i d_j}$
, where the arbitrary phases $d_j$ will be used in the fits to account for the duration of the switching process. The values for $c_j\in\mathbb{R}$ can be obtained from the height of the corresponding peak in the spectrum.

With this ansatz, expectation values like $\left<x^2\right>$  can be easily calculated, yielding a constant, time-independent part and a dynamic part made up of a sum over harmonic oscillations:
\begin{equation}
\left<x^2\right>(t) = \frac{1}{2} \sum_{j = 0}^m a_{j,j} + \sum_{j \neq k = 0}^m a_{j,k} \cos \left(\omega_{j,k} t + d_{j,k} \right)
\label{eqn:bvont}
\end{equation}
where the frequencies are given by $\omega_{j,k} = \omega_j - \omega_k$, the phases by $d_{j,k} = d_j - d_k$ and the amplitudes by
\begin{equation}
a_{j,k} = 2 c_k c_j \langle\Psi_k|x^2|\Psi_j\rangle.
\label{eqn:ajk}
\end{equation}


In an analogous way, also expectation values describing the center-of-mass dynamics like $\langle R \rangle$ or $\langle R^2 \rangle$ can be calculated. They share the frequencies of the modes $\omega_{j,k}$ with the dynamics of the relative motion but the amplitudes $a_{j,k}$ are different.

Within this few-mode model, we can now explain the dynamics of the two-boson system after excitation via distortion of the external trap. The frequencies $\omega_{j,k}$ of the modes contributing to the dynamics are determined by the frequencies of the contributing excited states. For weak excitations as mentioned above, only two excited states are considerably populated, and in this case it is expected that only three modes determine the dynamics of the system (for large interaction strength these are the ones with frequencies $\omega_{0,4}, \omega_{0,6}$ and $\omega_{4,6}$). Even more so, since 
the amplitude of the third mode (with frequency $\omega_{4,6}$) 
is much smaller than the amplitudes of the two other modes 
and has a much larger timescale ($\omega_{4,6} \ll \omega_{0,4}, \omega_{0,6}$), practically only two modes are expected to determine the dynamics of the two-particle system for weak excitations: $\omega^{(1)} = \omega_4 - \omega_0$ and $\omega^{(2)} = \omega_6 - \omega_0$. 
In this case we expect a simple beat behavior in the dynamics, where the beat period determining the time between two collapses is inversely proportional  to $\omega^{(2)}-\omega^{(1)} = \omega_6-\omega_4$ for large interaction strengths ($\omega_5-\omega_4$ for small ones). The condition for a complete collapse is the equality of the amplitudes of the two contributing modes ($a_{0,4} \approxeq a_{0,6}$), which is almost fulfilled in the relative dynamics but not valid for the center-of-mass breathing oscillations. So, by the way of excitation, e.g. the extent of anharmonicity added to the potential, it is possible to control the internal dynamics of the system, e.g., the time between two collapses of the oscillations.

\begin{figure}[t]
\includegraphics[width=.7\columnwidth]{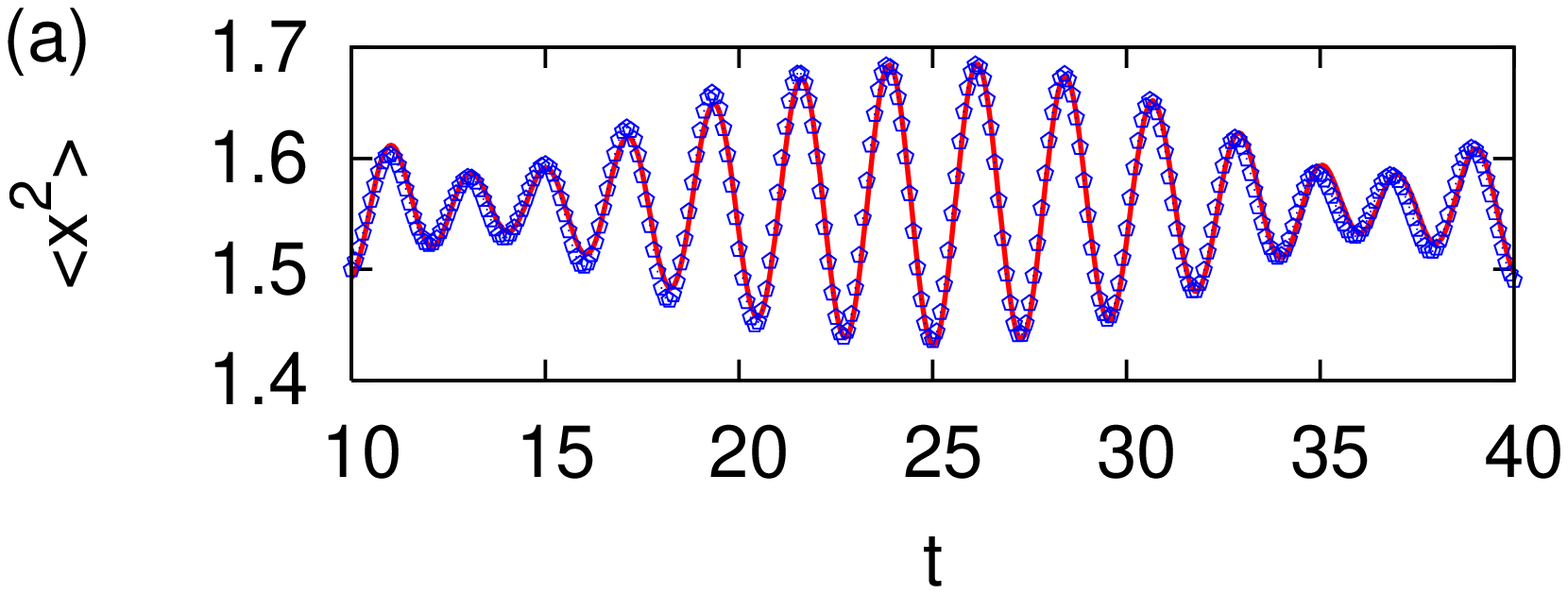} \\
\includegraphics[width=.7\columnwidth]{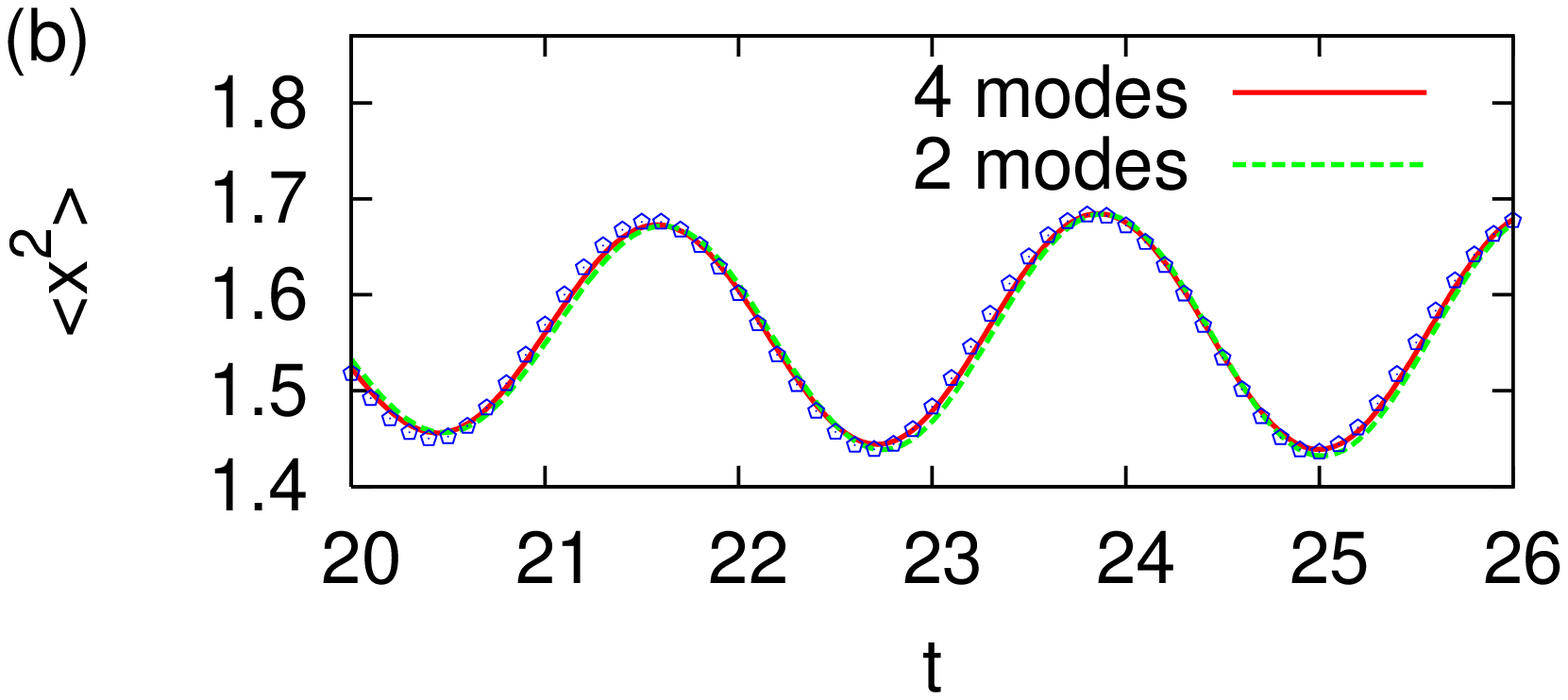} \\
\includegraphics[width=.7\columnwidth]{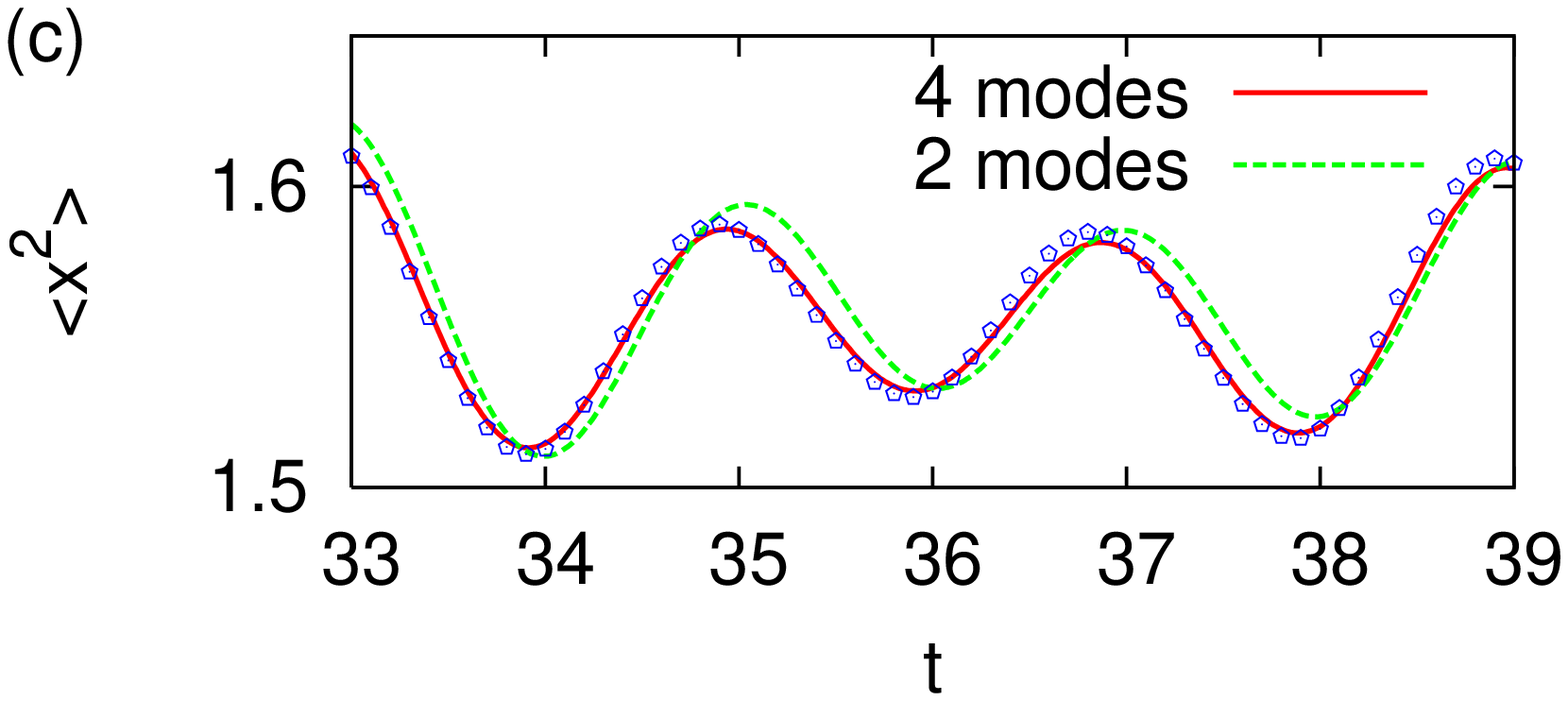} 

\caption{Fits of $\langle x^2 \rangle$ for a system with parameters $g = 20.0$ and $\lambda = 0.1$. The open pentagons  represent the exact simulation. (a) 2-mode fit for time interval $t \in [10,40]$ --- (b) and (c): more detailed 2-mode- and 4-mode fits, (b) for time interval $t \in [20,26]$, (c) for time interval $t \in [33,39]$.}
\label{fig:fits}
\end{figure}
 
Figure \ref{fig:fits} (a) confirms that the presented few-mode model (including 2 modes) fits the simulated data very well. Apart from that, the optimal fitted values for the frequencies of the modes coincide perfectly with the theoretically predicted values for $\omega_{0,4}$ and $\omega_{0,6}$. However, a closer look reveals that for regions in which the relative dynamics has a small amplitude [Fig. \ref{fig:fits}(b)], minor deviations emerge which can be fixed by including more modes in the theoretical model. Within an adequate 4-mode model (by adding the next strongest modes with frequencies  $\omega_{4,6}$ and  $\omega_{6,16}$), satisfactory agreement with the simulated data can be established. 
For stronger excitations, even more and higher states are considerably excited, and thus even more modes contribute significantly to the dynamics of the system. Then, of course, a pure 2-mode model no longer provides a good description for the dynamics of the system and for a satisfactory agreement between theory and simulation more modes must be included.

\subsubsection{Excitation without $Rx$-coupling term}

If the system is excited with an anharmonic potential in which artificially the $Rx$-coupling term is eliminated, it turns out that the same states are excited (i.e. the new ground state, the $4^{th}$, $6^{th}$ state and so on for larger interaction strengths) but some of the matrix elements emerging in equation (\ref{eqn:ajk}) vanish (e.g. $a_{0,4} = 0$) so that in the end fewer modes contribute to the dynamics of the system. 

For weak excitations, the dynamics is dominated by a single mode with frequency $\omega_{0,6}$ [cf. Fig. \ref{fig:collrev}(e)]; thus there is no beat behavior anymore. Hence, in contrast to the full excitation (including the $Rx$-coupling term in the Hamiltonian) there is no collapse in the relative dynamics. This gives a striking illustration of the role played by the coupling between the center-of-mass and the relative motion.

\subsection{Excitation by elongating the center of mass}
\label{sec:probl2}

\begin{figure}[t]
\includegraphics[width=.7\columnwidth]{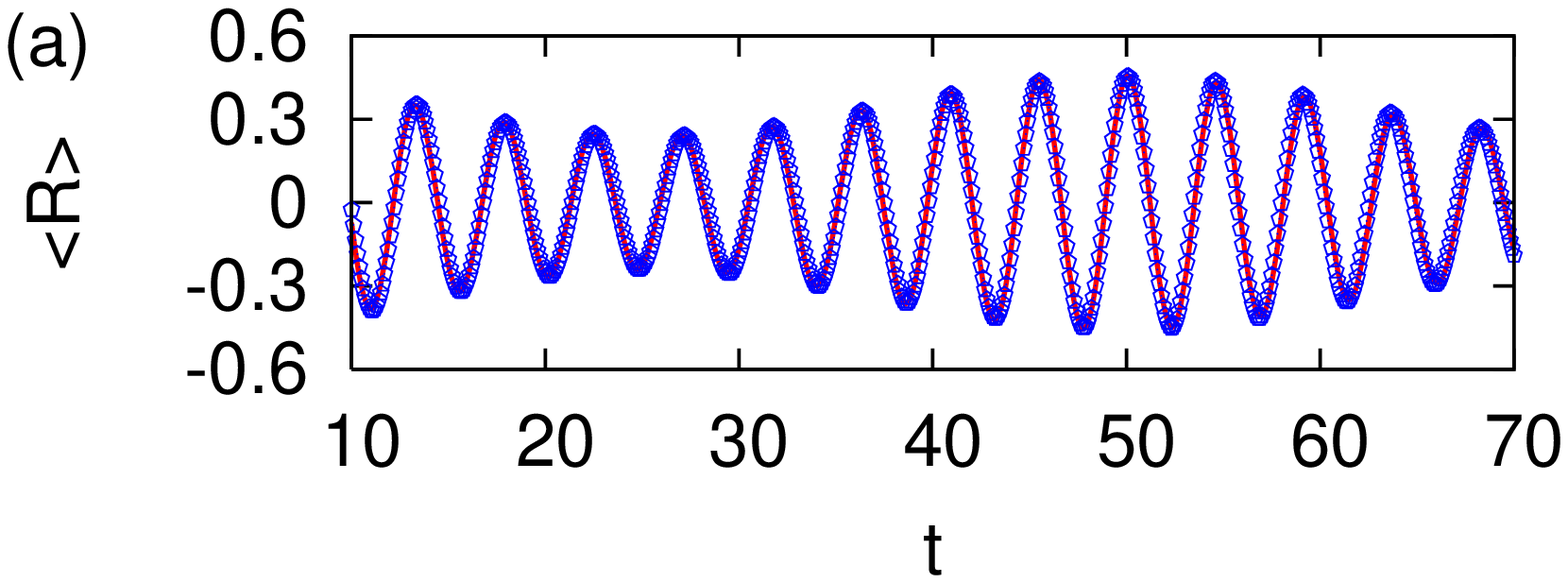} \\
\includegraphics[width=.7\columnwidth]{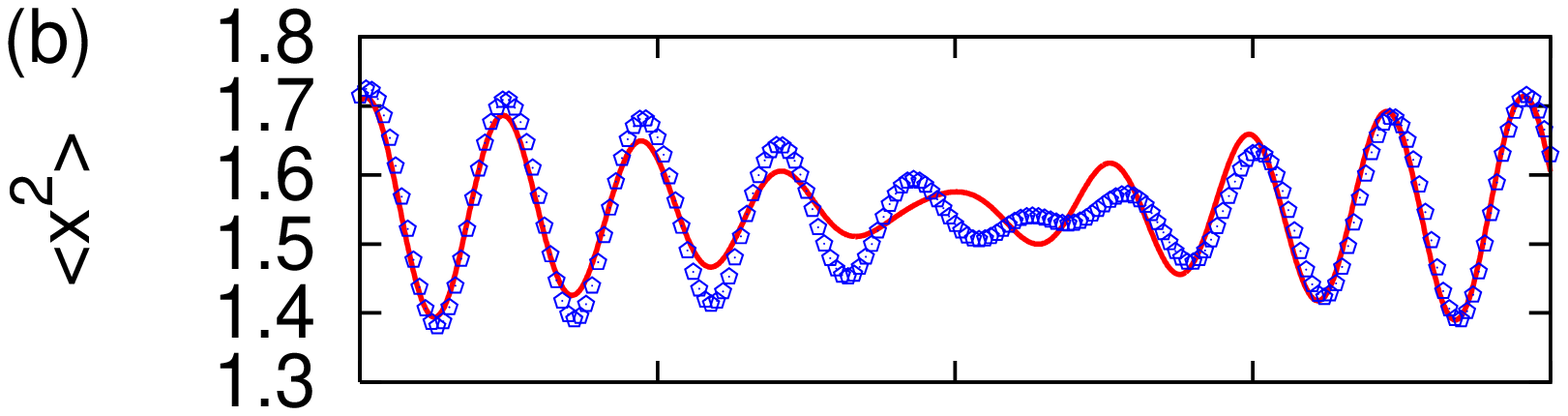} \\
\includegraphics[width=.7\columnwidth]{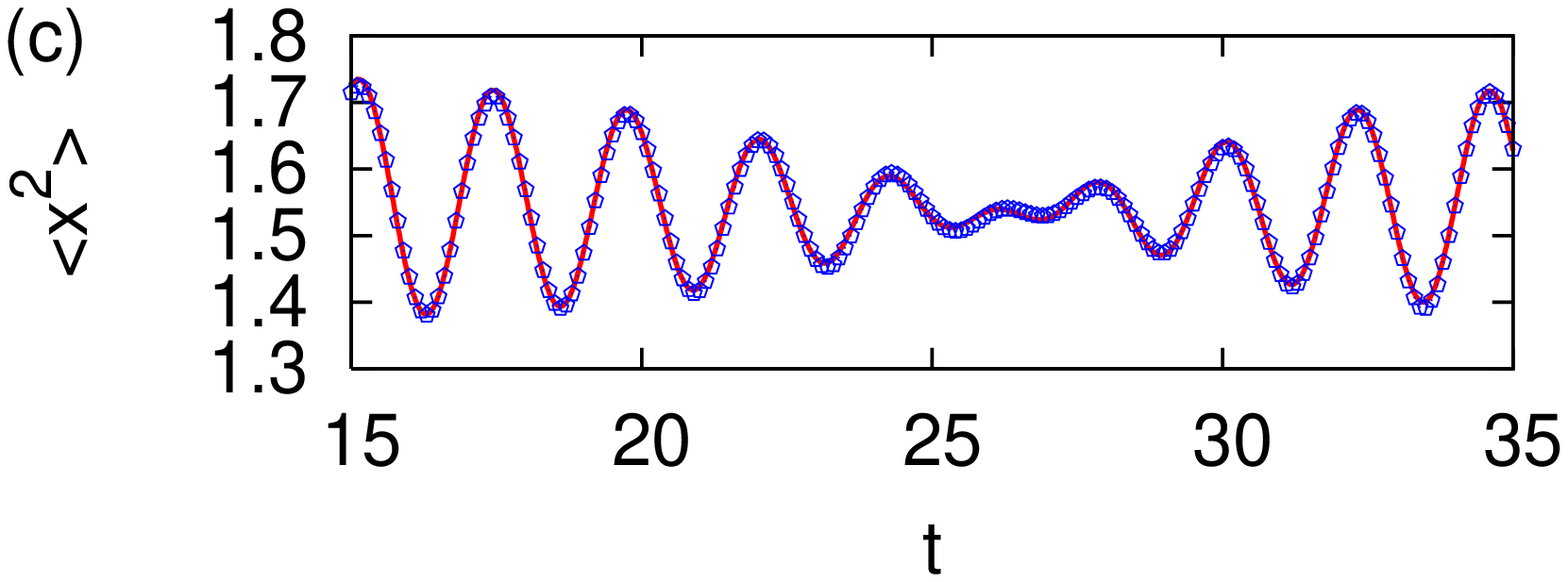} 

\caption{Dynamics in the anharmonic trap after elongating the center of mass. The open pentagons  represent the exact simulation. \\
(a) 3-mode fit of the center-of-mass position $\langle R \rangle$ for $g = 20.0$, $\lambda = 0.1$, $\kappa = 1.0$ \\
(b) and (c): Fits of the mean-square distance of the two particles $\langle x^2 \rangle$ for $g = 20.0$, $\lambda = 0.1$, $\kappa = 1.0$ in the time interval $t \in [15,35]$: (b) 3-mode fit, (c) 7-mode  fit.}
\label{fig:fits2}
\end{figure}

The second method of excitation examined in this paper uses the displacement of the trap center by a linear potential. We start with the ground state of the two-boson system in an anharmonic trap whose center is displaced (Eq.~(\ref{eqn:HaRr}) with $\lambda > 0, \kappa > 0$) and then continuously turn off the linear potential ($\kappa \rightarrow 0$) to revoke the displacement. 
With this procedure we excite collective oscillations: In fact, Fig. \ref{fig:fits2}(a) reveals how the center of mass oscillates about the trap center. Moreover, since the anharmonic force couples the center of mass to the relative coordinate, also breathing oscillations in the relative motion are induced. These are displayed in the time evolution of the widths $\Delta x^2 = \langle x^2 \rangle$ in Fig. \ref{fig:fits2}(b).
Again, we observe collapses and revivals of the oscillations in the relative dynamics, whereas in the center-of-mass motion the amplitude of the oscillations periodically changes with time but the oscillations never die out completely. As both excitation techniques are similar, we can adopt much of the argumentation from the last subsection and focus mainly on the differences. Altogether, we find that the dynamics is more complicated than in the previous scheme. The reason for that will be laid out in the following two paragraphs.




\subsubsection{Excitation spectra and symmetry analysis}

An analysis of the spectra shows that, on top of the states already excited with the previous excitation scheme, now more states contribute, the lowest ones being the $2^{nd}$, $8^{th}$ and $10^{th}$ state.
However, along the same lines as before, this is readily explained in terms of the displacement term $\kappa R$ in the Hamiltonian (\ref{eqn:HaRr}), which breaks the $R$-parity (so that only the trivial $x$-parity is conserved during the excitation process). Hence, the $2^{nd}$, $8^{th}$ and $10^{th}$ state ---all having odd $R$-parity---are now also available.

\subsubsection{Few-mode model}

Equipped with this knowledge, the simulation can again be explained within the simple few-mode model introduced in Sec.~\ref{sec:fewmode}. The only thing is that, as now more states are excited, also more modes significantly contribute to the dynamics, which is therefore more complicated as compared to the other excitation method. Especially the $2^{nd}$ state is, next to the ground state, the most populated state after excitation, but also the $8^{th}$ and $10^{th}$ state are more populated than e.g. the $16^{th}$. As an example, for an interaction strength of $g = 20.0$, an anharmonicity of $\lambda = 0.1$ and a displacement of $\kappa = 1.0$, Figs. \ref{fig:fits2} (b) and (c) illustrate that for the relative dynamics at least seven modes are necessary to have a satisfactory agreement of the simulation with the theoretical model. On the other hand, the dynamics of the center of mass can already be very well described in terms of only three modes having frequencies $\omega_{0,2}, \omega_{2,4}$ and $\omega_{2,6}$, see Fig. \ref{fig:fits2}(a).

\section{Conclusion and outlook}

In this work we have investigated a repulsively interacting two-boson system in a one-dimensional anharmonic trap. We have investigated both the ground state of the system as well as its quantum dynamics upon excitation via distortion of the trap, with an eye toward the impact of the coupling between center of mass and relative motion.
Our calculations are based on the numerically exact Multi-Configuration Time-Dependent Hartree method. 

Two different methods of excitation have been applied. In the first one we continuously switch on an anharmonic potential in addition to an otherwise harmonic trap. This process induces breathing oscillations in the relative motion, which experience collapses and revivals. The dynamics can be explained within a simple few-mode model based on the contributing excited states. For weak excitations, only two modes determine the dynamics of the system, while for stronger excitations more modes must be included. Just which states are excited depends on the symmetries of the initial state and thus on the method of excitation. With this knowledge it is possible to control the internal motion of the system---viz., the amplitude of the oscillations or the time between two collapses---by the external excitation via an adequate distortion of the trapping potential. 
We have also illuminated the role played by the coupling between the center of mass and the relative motion by artificially excluding the corresponding terms in the Hamiltonian. Then, at least in the limit of weak excitations, one single mode dominates the oscillations, which are thus etirely undamped. 


In the second excitation scheme, we trigger collective oscillations by displacing the center of mass, whose coupling to the relative coordinate leads to similar internal excitations as before. However, since the initial displacement breaks the center-of-mass parity, the dynamics is now more complex.

An obvious extension of our work would be the study of the dynamics of systems with more than two bosons, this way gaining insight into anharmonicity effects on a many-body level. One could also conceive more sophisticated excitation schemes so as to get detailed probes of the internal excitation. This may help not only understand experimental effects beyond harmonic confinement, but also to actively control the interatomic dynamics.
 


\begin{acknowledgments}
Financial support by the Landesstiftung Baden-Württemberg in the framework
of the project ``Mesoscopics and atom optics of small ensembles
of ultracold atoms'' is gratefully acknowledged by P.S. and S.Z. 
\end{acknowledgments}

\bibliographystyle{prsty}
\bibliography{/home/christian/bib/literature,/home/sascha/bib/phd,/home/sascha/bib/mctdh}

\end{document}